# Comment on "Demographic gaps or preparation gaps?: The large impact of incoming preparation on performance of students in introductory physics"

M. B. Weissman

*Department of Physics, University of Illinois at Urbana-Champaign*

1110 West Green Street, Urbana, IL 61801-3080

A recent paper by Salehi et al. [1] claims that the differences found between major demographic groups on scores in introductory college physics tests are due to differences in pre-college "preparation". No evidence is produced, however, to show that preparation differences are more causally important than any other differences. In one case, the male/female difference, the paper actually provides evidence that preparation gaps are unimportant.

The explicit data analysis given shows that the college exam scores of individuals can be predicted fairly well from a combination of pre-college scores on math ACT/SAT tests and the physics "concept inventory" (CI) exams. [1] If demographic variables are added to the predictive model, their coefficients are small for practical purposes and also not large enough to consider statistically significant in this sample. [1] This result is then interpreted to mean that "preparation gaps" are responsible for the demographic differences in college scores.

The initial test differences, like the final test differences, can arise from any number of common causes, including ones for which measures are not available. Any fairly stable individual trait measurable by these exams would give the results observed, so long as it had about the same effect on the initial and final tests and so long as there were not large demographic differences whose effects differed on the two test sets. Thus the core result, the insignificance of direct demographic prediction terms for the college tests once pre-college tests are included in the model, does not provide information to distinguish what causes the stable individual differences. To some extent, the paper acknowledges that by saying that structural equation modeling analysis "does not test for causality", but the ordinary language interpretation of the title and the entire discussion implies that "preparation" in something like the usual sense of the word is the key causal factor. The Conclusion extensively discusses the implications for how courses should be changed in order to change outcomes, i.e. it makes the counterfactual causal interpretation very explicit.

The key sentence of the paper comes in the Discussion: "We initially expected that it would be differences in what high school physics courses were taken, but we analyzed that for HSWCU [the highly selective west-coast university] , and we found that all demographic groups at this institution had the same distribution of taking AP physics, regular high school physics, and no physics, even though the groups had different average CI prescores and math SAT or ACT scores." In other words, the main conventional component of "preparation" was at least nominally the same for the different groups. That would suggest that preparation is not the key variable that differs.

Realistically, however, courses with the same name can be radically different in different US schools, and those differences are likely to show major correlations with racial/ethnic differences. Therefore the results do not provide strong evidence against the role of preparation in causing differences between those groups, although they provide no evidence for the role of preparation. For the most part, however, males and females go to the same schools, so that if they took the same nominal courses they took the same *actual* courses. Thus the results provide strong evidence that it is *not* preparation, at least in the ordinary sense of the word, that accounts for the male/female differences in both precollege and college exams. One could broaden the definition of "preparation" to include all causes, including unmeasured or unmeasurable ones, that affect prior test scores, but that would make the claim that preparation is the key difference an untestable proposition.

Policy recommendations in social sciences, including education, rely on estimating causal effects. The notorious difficulties in drawing reliable causal conclusions from observational data in these fields should be taken as a reason to pay even closer attention to the relation between model predictions and data than we do in physics, rather than as a reason to relax that attention.

I thank Carl Wieman for a cordial exchange and Jamie Robins and Sander Greenland for very helpful editorial comments.